\documentclass[reprint, superscriptaddress,  aps]{revtex4-2}
\usepackage{amsfonts}
\usepackage{amssymb}
\usepackage{amsmath}
\usepackage{graphics}
\usepackage{graphicx}
\usepackage{subfigure}
\usepackage{dcolumn}
\usepackage{bm}
\usepackage{color}
\usepackage{epstopdf}

\definecolor{Gray}{gray}{0.1}
\DeclareMathAlphabet\mathbfcal{OMS}{cmsy}{b}{n}

\begin{document}

\title{Berry curvature and shift vector effects at high-order wave mixing in
biased bilayer graphene}
\author{H.K. Avetissian}
\affiliation{Centre of Strong Fields Physics at Research Institute of Physics, Yerevan State University,
Yerevan 0025, Armenia}

\author{H.H. Matevosyan}
\affiliation{Institute of Radiophysics and Electronics NAS RA, Ashtarak 0203, Armenia}

\author{G.F. Mkrtchian}
\thanks{mkrtchian@ysu.am}
\affiliation{Centre of Strong Fields Physics at Research Institute of Physics, Yerevan State University,
Yerevan 0025, Armenia}

\begin{abstract}
In this work, we present a microscopic quantum theory that elucidates the nonlinear and nonperturbative optical response of biased bilayer graphene subjected to bichromatic strong laser fields. This response is analyzed using a four-band Hamiltonian derived from \textit{ab initio} calculations. For the laser-stimulated dynamics, we employ structure gauge-invariant evolutionary equations to accurately describe the evolution of the single-particle density matrix across the entire Brillouin zone. The resonant generation of electron-hole pairs by the high-frequency component of the field, combined with the induction of high-order harmonic generation and high-order wave mixing by the strong low-frequency field component, leads to significant alterations in the resulting spectra. These changes are driven by the effects of Berry curvature and the shift vector, which modify the relative contributions of interband and intraband channels, thereby fundamentally reshaping the radiation spectra at high-order frequency multiplication. The numerical results are further supported by approximate analytical calculations, demonstrating that high-order wave mixing can be modeled using the classical trajectory analysis of electron-hole pairs, with Berry curvature and the shift vector significantly influencing the saddle-point equations.

\end{abstract}

\maketitle

\section{Introduction}

In recent decades, there has been a growing interest in extending atomic
High Harmonic Generation (HHG) and High-Order Wave Mixing (HWM) phenomena to
novel nanostructures \cite{ghimire2019high,li2020attosecond,brabec2022high}.
Beyond its inherent interest as a frequency multiplication process and its
capability to generate high-energy photons \cite{avetissian2015relativistic}%
, HHG also represents a promising tool for various applications in
condensed-matter systems. These applications include probing the structure 
\cite{vampa2015all,luu2015extreme,lakhotia2020laser}, exploring the topology
of bands \cite%
{kelardeh2017graphene,bauer2018high,silva2019topological,chacon2020circular,schmid2021tunable,bai2021high}%
, and investigating electron dynamics/correlation effects \cite%
{vampa2015linking,tancogne2017impact,silva2018high,murakami2021high,neufeld2021light}%
, among others.

Advancements in nanotechnology over recent decades have enabled the
synthesis of a new class of optical materials, namely two-dimensional
nanostructures with extremely high carrier mobility and extraordinary
properties \cite{geim2013van}. Among these materials, gapped graphene-like
systems with non-trivial topology of energy bands are of particular
interest. Examples include atomically thin monolayer transition metal
dichalcogenides \cite{manzeli20172d} and monolayer hexagonal boron nitride 
\cite{caldwell2019photonics}. Technological advances now also allow the
realization of bilayers of these materials, where the stacking order and
twist angle between the layers can significantly alter the nanostructure's
electronic and optical properties, giving rise to superconductivity,
interaction-induced insulating states, magnetism, and quantized anomalous
Hall states \cite{andrei2020graphene}.

Recently, there has been increasing interest in investigating the nonlinear
electromagnetic response of crystalline systems with a non-vanishing Berry
curvature. In these systems, a laser field can induce a nonlinear anomalous
current polarized perpendicular to the laser polarization, resulting in the
generation of non-perturbative anomalous high harmonics \cite%
{liu2017high,luu2018measurement,lv2021high,avetissian2020high,dantas2021nonperturbative,bharti2022high,avetissian2022high,yue2023characterizing,bharti2023role,medic2024high}%
. The shift vector, which is related to the polarization of the bands during
direct optical transitions, also plays a significant role \cite%
{qian2022role,parks2023gauge} and can be utilized for investigating
topological phase transitions \cite{qian2022role}.

Bilayer graphene (BLG) with Bernal stacking is the most commonly realized
form in experiments \cite%
{ohta2006controlling,mccann2013electronic,rozhkov2016electronic}, and its
electronic properties are considerably richer than those of monolayer
graphene \cite{mccann2006landau,guinea2006electronic,koshino2006transport}.
In BLG, the interlayer coupling between the two graphene sheets alters the
monolayer's Dirac cone, inducing trigonal warping in the band dispersion and
changing the topology of the Fermi surface in the low-energy region \cite%
{mccann2013electronic}. The nonlinear optical response of BLG has been
extensively investigated in recent literature \cite%
{avetissian2013multiphoton,yang2014giant,ulstrup2014ultrafast,mcgouran2016nonlinear,mcgouran2017nonlinear,hipolito2018nonlinear,avetissian2020high2, ghazaryan2020high,mrudul2021high,ren2022orientation}%
. Specifically, HHG from BLG near the Dirac points has been explored using
low-energy Hamiltonians in works \cite%
{avetissian2013multiphoton,avetissian2020high2, ghazaryan2020high}.
High-frequency scenarios, utilizing phenomenological tight-binding
Hamiltonians applicable to the full Brillouin zone (BZ), have been
considered in more recent studies \cite{mrudul2021high,ren2022orientation}.
Additionally, there is a growing interest in examining the nonlinear optical
response and HHG from twisted bilayer graphene, as highlighted in several
recent publications \cite%
{topp2019topological,ikeda2020high,du2021high,zuber2021nonlinear,di2022optotwistronics,molinero2024high}%
.

An additional advantage of bilayer materials is that applying a
symmetry-lowering perpendicular electric field can control or induce a band
gap \cite{guinea2006electronic,aoki2007dependence}, thereby modifying the
topology of the bands. Thus, biased bilayer graphene with Bernal stacking
exhibits nontrivial Berry curvatures in its energy bands \cite{xiao2010berry}%
. Along with the Berry curvatures, the shift vector is ubiquitous in modern
condensed-matter theory and is responsible for various topological effects 
\cite{xiao2010berry,vanderbilt2018berry}. The tunable band gap in BLG can be
adjusted up to approximately 250 meV \cite{tang2011electric}, which is of
practical importance. Furthermore, the Berry curvature near the band edge in
biased bilayer graphene is about two orders of magnitude larger than that in
monolayer hexagonal boron nitride and MoS$_{2}$. This unique feature
provides an unprecedented opportunity to investigate the influence of Berry
curvature and shift vector on the HHG process.

An effective approach to enhancing HHG is the use of multicolor driving
pulses. High-order wave mixing processes with multicolor laser fields have
garnered significant interest due to the additional degrees of freedom they
offer, such as relative polarizations, intensities, phases, and wavelengths
of the involved pump waves. HHG with different compositions of driving laser
pulses has been studied in crystalline systems, considering two distinct
regimes: first, when the driving field consists of the fundamental wave and
its harmonics \cite%
{li2017enhancement,luu2018observing,shirai2018high,shao2020quantum,song2020enhanced,navarrete2020two,mrudul2021light,avetissian2022efficient}%
; and second, when one of the involved wave frequencies is significantly
higher than the other \cite%
{yan2008theory,zaks2012experimental,xie2013effects,crosse2014quantum,langer2016lightwave,langer2018lightwave,avetissian2019wave,avetissian2020many}%
. However, high-order wave mixing in bilayer graphene and a thorough
investigation of the role of Berry curvature and shift vector remain
unexplored.

In this work, we investigate the non-perturbative optical response of biased
bilayer graphene to a two-color strong laser field, employing a four-band
Hamiltonian derived through \textit{ab initio} calculations. For the laser-stimulated
dynamics, we employ structure gauge-invariant evolutionary equations \cite%
{parks2023gauge} to accurately describe the evolution of the single-particle
density matrix across the entire BZ. We consider the efficient generation of
electron-hole pairs by the high-frequency component of the field, coupled
with acceleration by the low-frequency strong field component, which leads
to a significant alteration of the HWM and HHG spectra due to the Berry
curvature. This paper is organized as follows. In Sec. II, the theoretical
model and numerical methods are presented. In Sec. III, we present the main
results. Finally, conclusions are given in Sec. IV.

\section{The model and theoretical approach}

We begin by describing the model and theoretical approach. Let a BLG nanostructure interact 
with two-color plane electromagnetic radiation. We consider the interaction scenario 
where the waves propagate in a direction perpendicular to the nanostructure (in the $XY$ plane). 
The schematic representation of this interaction is shown in Fig. 1. In this configuration, these traveling waves become homogeneous quasiperiodic electric fields with carrier frequencies $\omega _{1}$ and $\omega _{2}$. We assume that the waves are linearly polarized along the $x$
direction:
\begin{equation}
\mathbf{E}\left( t\right) =\widehat{\mathbf{x}}\left( f_{1}\left( t\right)
E_{01}\cos \omega _{1}t+f_{2}\left( t\right) E_{02}\cos \omega _{2}t\right) .
\label{E_field}
\end{equation}
\begin{figure}[tbp]
\includegraphics[width=0.33\textwidth]{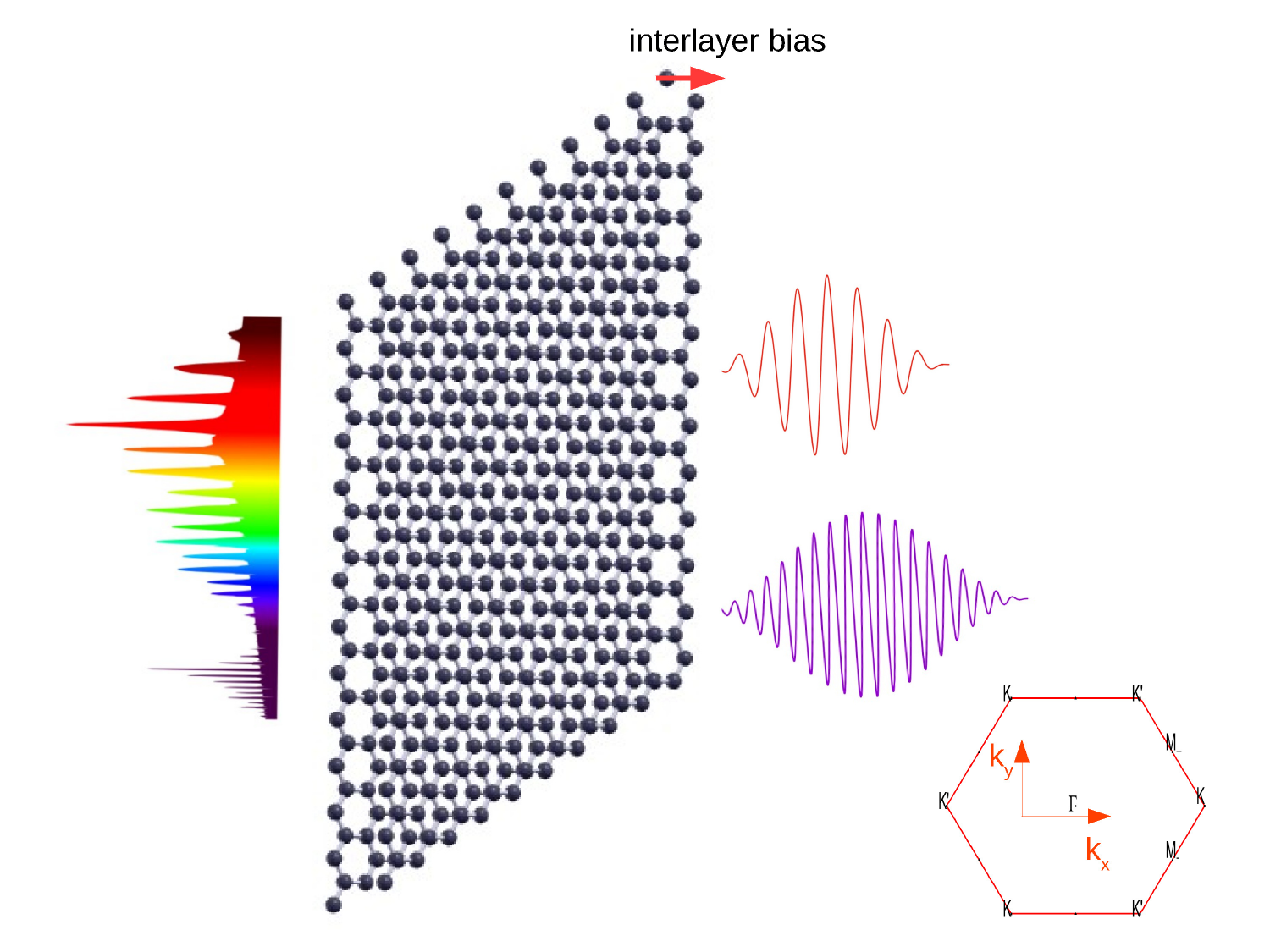}
\caption{The schematic representation of interaction and the first BZ of the
reciprocal lattice with high-symmetry points.}
\end{figure}
The wave envelopes are described by the sin-squared function $f_{1,2}\left(
t\right) =\sin ^{2}\left( \pi t/\mathcal{\tau }_{1,2}\right) $, where $%
\mathcal{\tau }_{1}$ and $\mathcal{\tau }_{2}$ are the durations of the wave
pulses. We performed our calculations within the structure-gauge\textit{\ }%
invariant formalism \cite{parks2023gauge}, formulated for the
gauge-invariant coherence matrix $\varrho _{nm}(\mathbf{k},t)$, whose
diagonal elements represent the band occupations. We denote these as $%
N_{n}(\mathbf{k},t)\equiv \varrho _{nn}(\mathbf{k},t)$. For the sake of compactness of the equations, the elementary charge 
and Planck's constant are set to 1 throughout the paper, unless stated otherwise. The upper indices denote Cartesian
coordinates. The structure-gauge invariant semiconductor Bloch equations (GI-SBEs) 
\cite{parks2023gauge} in the length gauge for band occupations read 
\begin{equation}
\begin{aligned}
&\partial _{t}N_{n}(\mathbf{k},t)-E(t)\partial _{k^{x}}N%
_{n}(\mathbf{k},t)\\
&=2E(t)\sum\limits_{l\neq n}\left\vert r_{nl}^{x}(\mathbf{k}%
)\right\vert \mathrm{Im}\varrho _{ln}(\mathbf{k},t)  \label{SBE1}
\end{aligned}
\end{equation}%
and for interband coherence:%
\begin{equation}
\begin{aligned}
&\partial _{t}\varrho _{nm}(\mathbf{k},t)-E(t)\partial _{k^{x}}\varrho _{nm}(%
\mathbf{k},t)=-\Gamma \varrho _{nm}(\mathbf{k},t) \\
&-i\left( \mathcal{E}_{nm}\left( \mathbf{k}\right) +E(t)R_{nm}^{x,x}\left( 
\mathbf{k}\right) \right) \varrho _{nm}(\mathbf{k},t) \\
&+i\left\vert r_{nm}^{x}(\mathbf{k})\right\vert \left( N_{n}(%
\mathbf{k},t)-N_{m}(\mathbf{k},t)\right) E(t) \\
&+i\sum\limits_{l\neq n,m}T_{nlm}\left( \left\vert r_{lm}^{x}(\mathbf{k}%
)\right\vert \varrho _{nl}(\mathbf{k},t)-\left\vert r_{nl}^{x}(\mathbf{k}%
)\right\vert \varrho _{lm}(\mathbf{k},t)\right) E(t),  \label{SBE2}
\end{aligned}
\end{equation}%
where 
\begin{equation}
r_{nm}^{b}(\mathbf{k})=\langle n,\mathbf{k}|i\partial _{k^{b}}|m,\mathbf{k}%
\rangle =e^{i\Phi _{nm}^{b}(\mathbf{k})}\left\vert r_{nm}^{b}(\mathbf{k}%
)\right\vert  \label{tdm}
\end{equation}%
is the transition dipole moment (TDM) between bands $m$ and $n$, 
\begin{equation}
R_{nm}^{b,a}(\mathbf{k})=-\partial _{k^{b}}\Phi _{nm}^{a}(\mathbf{k})+%
\mathcal{A}_{\mathrm{n}}^{b}\left( \mathbf{k}\right) -\mathcal{A}_{\mathrm{m}%
}^{b}\left( \mathbf{k}\right)  \label{shift}
\end{equation}%
is the gauge invariant quantity known as the shift vector \cite%
{sipe2000second}, 
\begin{equation}
\mathcal{A}_{\mathrm{n}}^{b}\left( \mathbf{k}\right) =\langle n,\mathbf{k}%
|i\partial _{k^{b}}|n,\mathbf{k}\rangle  \label{bc}
\end{equation}%
is the Berry connection for band $n$, $\mathcal{E}_{nm}\left( \mathbf{k}%
\right) =\mathcal{E}_{n}\left( \mathbf{k}\right) -\mathcal{E}_{m}\left( 
\mathbf{k}\right) $ is the energy difference between bands, $T_{nlm}(\mathbf{%
k})=e^{i\left( \Phi _{lm}^{x}(\mathbf{k})+\Phi _{nl}^{x}(\mathbf{k})+\Phi
_{mn}^{x}(\mathbf{k})\right) }$ is the gauge invariant triple phase product 
\cite{parks2023gauge}, and $\Gamma $ describes relaxation processes. The
single particle density matrix $\rho _{nm}\left( \mathbf{k},t\right) $ is
defined via the gauge invariant coherence matrix $\varrho _{nm}(\mathbf{k}%
,t) $ as follows: 
\begin{equation}
\rho _{nm}\left( \mathbf{k},t\right) =\varrho _{nm}(\mathbf{k},t)e^{i\Phi
_{nm}^{x}(\mathbf{k})}.  \label{density}
\end{equation}%
The Berry curvature for band $m$ is defined as the curl of the Berry
connection: 
\begin{equation}
\bm{\mathcal{B}}_{m}\left( \mathbf{k}\right) =\partial_{\mathbf{k}}\times\bm{\mathcal{A}}_{m}\left( \mathbf{k}\right). \label{Berry_main}
\end{equation}
It is important to note that all quantities in Eqs. (\ref{SBE1}) and (\ref%
{SBE2}) are structure-gauge invariant.

The matrix elements required for the GI-SBEs are obtained numerically. For
this purpose, we performed an ab-initio calculation using the HSE06
functional and a 12x12 Monkhorst-Pack grid with the QuantumEspresso code 
\cite{giannozzi2009quantum}. Following this, we projected onto the $p_{z}$
orbitals and employed a wannierization procedure \cite{mostofi2008wannier90}
to obtain the 4x4 Hamiltonian $\widehat{H}_{nm}$. An interlayer bias is
incorporated via diagonal Hamiltonian $\widehat{H}_{bias}$ \cite%
{yang2014giant}. We then numerically diagonalized the Hamiltonian $\widehat{H%
}_{nm}+\widehat{H}_{bias}$, obtaining the eigenvectors $|\alpha ,\mathbf{k}%
\rangle $ and eigenvalues. The TDM magnitudes and triple products were
derived from the gradient of the Hamiltonian, while the shift vectors and
Berry curvatures were determined using a generalized Wilson loop algorithm 
\cite{wang2022generalized}. The Wilson loop can be defined as%
\begin{equation*}
\begin{aligned}
W_{\beta \alpha }^{a,b}\left( \mathbf{k,q}_{a}\right) &=\langle \beta ,%
\mathbf{k}||\beta ,\mathbf{k+q}_{a}\rangle \langle \beta ,\mathbf{k+q}%
_{a}|r^{a}|\alpha ,\mathbf{k+q}_{a}\rangle \\
&\times \langle \alpha ,\mathbf{k+q}_{a}||\alpha ,\mathbf{k}\rangle \langle
\alpha ,\mathbf{k}|r^{b}|\beta ,\mathbf{k}\rangle .
\end{aligned}
\end{equation*}%
We then arrive at the Wilson loop representation of the Berry curvature
tensor%
\begin{equation}
\Omega _{mn}^{ab}\left( \mathbf{k}\right) =2\mathrm{Im}\left[
W_{mn}^{a,b}\left( \mathbf{k,q}_{a}=0\right) \right] ,  \label{berry}
\end{equation}%
and the shift vector%
\begin{equation}
R_{mn}^{a,b}\left( \mathbf{k}\right) =-\mathrm{lim}_{q_{a}\rightarrow
0}\partial _{q_{a}}\mathrm{arg}\left[ W_{mn}^{b,b}\left( \mathbf{k,q}%
_{a}\right) \right] .  \label{shiftV}
\end{equation}%
Notably, these calculations of gauge-invariant quantities do not require the
use of a smooth Bloch gauge. Using these quantities, the Berry curvature for
band $m$ can be computed \cite%
{wang2022generalized,wang2006ab,wang2019ferroicit}. Since the system under
consideration is two-dimensional, it is particularly meaningful to discuss
the z-component of the Berry curvature. The local Berry curvature between
bands $m$ and $n$ is $\mathcal{B}_{mn}^{z}\left( \mathbf{k}\right) =\Omega _{mn}^{xy}\left( 
\mathbf{k}\right)$,
while the global Berry curvature for band $m$ is expressed as: 
\begin{equation}
\mathcal{B}_{m}^{z}\left( \mathbf{k}\right) =\sum_{n\neq m}\mathcal{B}%
_{mn}^{z}\left( \mathbf{k}\right) .  \label{Berry1}
\end{equation}%
The global Berry curvature defined above is equivalent to the expression in
Eq. (\ref{Berry_main}). Additionally, it is useful to note the relationship
between two key topological quantities, the shift vector and the Berry
curvature:%
\begin{equation}
\mathcal{B}_{n}^{z}-\mathcal{B}_{m}^{z}=\partial
_{k^{x}}R_{nm}^{y,a}-\partial _{k^{y}}R_{nm}^{x,a},\ a=x,y.  \label{Berry2}
\end{equation}%
This relationship follows directly from Eqs. (\ref{shift}) and (\ref%
{Berry_main}), and it establishes that the curl of the shift vector
corresponds to the difference in the Berry curvatures of two bands.
\begin{figure}[tbp]
\includegraphics[width=0.39\textwidth]{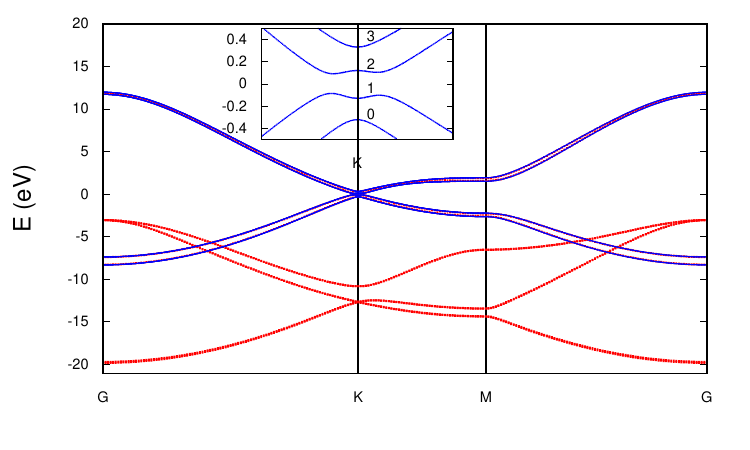}
\caption{The electronic band structure calculated using the Quantum Espresso
code is shown in red (dots) and projected onto the $p_{z}$ orbitals in blue
(solid line). The inset displays a magnified view of the band structure near
the K point, with corresponding band labels. The interlayer bias is 250 meV.}
\end{figure}
In Fig. 2 we show the electronic band structure calculated using the Quantum
Espresso code, with projection onto the $p_{z}$ orbitals via a
Wannierization procedure. Considering only the $p_{z}$ orbitals, BLG
features two conduction bands (labeled 2 and 3) and two valence bands (0 and
1). We assume the Fermi level is centered within the energy gap, with the
system initially in its ground state, where the conduction bands are empty
and the valence bands are fully occupied. The inset provides a magnified
view of the band structure near the $K$ ($K^{\prime }$) valley showcasing
the well-known "Mexican hat" structure. Figure 3 illustrates the Berry
curvature for band second band calculated via Eq. (\ref{Berry1}). As seen,
the Berry curvature is predominantly concentrated at the corners of the 2D
hexagonal BZ, particularly near the Dirac points. Due to time-reversal (TR)
symmetry the $K$ and $K^{\prime }$ valleys exhibit Berry curvatures with
opposite signs, with substantial values near the band edges. It is about two
orders of magnitude larger than that in monolayer MoS$_{2}$.
\begin{figure}[tbp]
\includegraphics[width=0.36\textwidth]{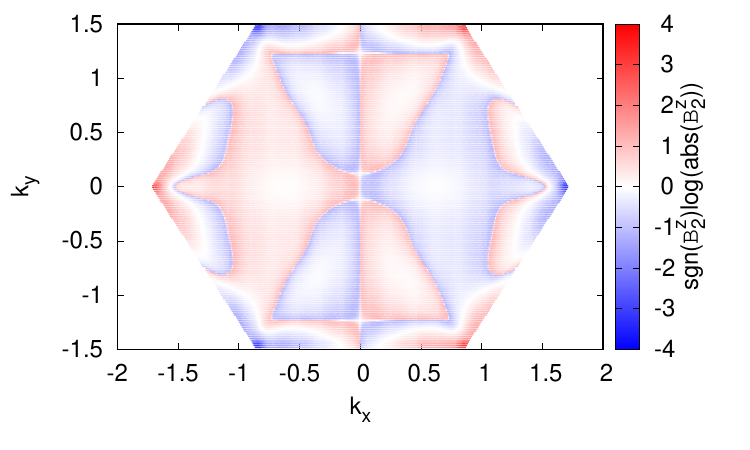}
\caption{A density plot of the Berry curvature for band 2, presented on a
logarithmic scale.}
\end{figure}
Figure 4 depicts the shift vector $R_{12}^{x,x}\left( \mathbf{k}\right) $
for the transition from the top of the valence band to the bottom of the
conduction band, calculated via Eq. (\ref{shiftV}). As evident from the
figure, the shift vector exhibits significant values across the 2D hexagonal
BZ, especially near the zone edges and high-symmetry points. Due to TR
symmetry, the shift vector changes sign under the transformation $\mathbf{%
k\rightarrow -k}$. The shift vector's geometric interpretation relates to
the difference in real-space charge centers (or polarization) between bands
during a direct optical transition, which is expected to have a significant
impact on the HHG and HWM spectra.
\begin{figure}[tbp]
\includegraphics[width=0.36\textwidth]{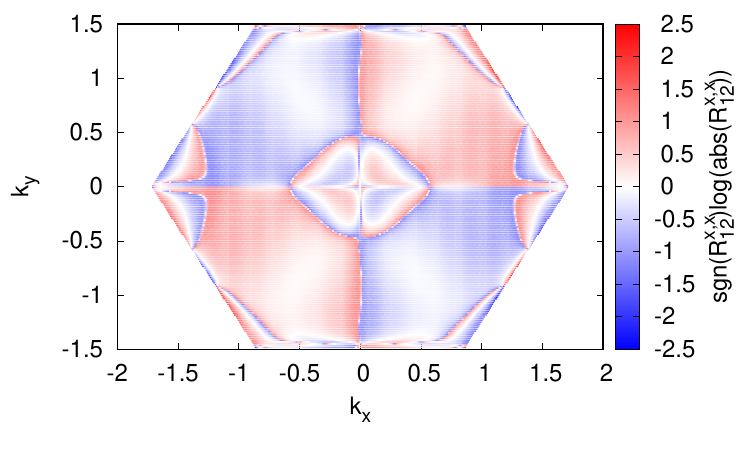}
\caption{A density plot of the shift vector for the transition from the top
of the valence band to the bottom of the conduction band, shown on a
logarithmic scale.}
\end{figure}
Optical excitation induces a surface current that can be calculated using
the following formula:
\begin{equation}
\begin{aligned}
j^{a}\left( t\right) &=-\sum\limits_{n\mathbf{k}}\partial _{k^{a}}\mathcal{E}%
_{n}(\mathbf{k)}\rho _{nn}(\mathbf{k},t) \\
&-i\sum\limits_{n\neq m}\sum\limits_{\mathbf{k}}r_{mn}^{a}\left( \mathbf{k}%
\right) \mathcal{E}_{mn}\left( \mathbf{k}\right) \rho _{nm}(\mathbf{k},t)
\label{tcur}
\end{aligned}
\end{equation}%
The latter can be written with the help of coherence matrix as follow 
\begin{equation}
\begin{aligned}
&j^{a}\left( t\right) =-\sum\limits_{n\mathbf{k}}\partial _{k^{a}}\mathcal{E}%
_{n}(\mathbf{k)}\varrho _{nn}(\mathbf{k},t) \\
&-i\sum\limits_{n\neq m}\sum\limits_{\mathbf{k}}\left\vert r_{nm}^{a}(\mathbf{%
k})\right\vert \mathcal{E}_{mn}\left( \mathbf{k}\right) \varrho _{nm}(%
\mathbf{k},t)e^{i\left( \Phi _{nm}^{x}-\Phi _{nm}^{a}\right) }
\label{tcurinv}
\end{aligned}
\end{equation}%
The first term in Eq. (\ref{tcurinv}) is due to intraband transitions, while
the second term is due to interband transitions. As is clear second term
includes also topological part of current. To understand how the non-trivial
topology of biased BLG relates to the generated current, we derive an
equivalent equation for the current (\ref{tcurinv}) that explicitly includes
both Berry curvature and the shift vector. By substituting the expression
for $i\mathcal{E}_{mn}\left( \mathbf{k}\right) \varrho _{nm}(\mathbf{k},t)$
from Eq. (\ref{SBE2}) into the equation for the current (\ref{tcurinv}),
taking into account the sum roles \cite{sipe2000second}%
\begin{equation*}
\begin{aligned}
&\partial _{k^{b}}\mathcal{A}_{\mathrm{\alpha }}^{a}\left( \mathbf{k}\right)
-\partial _{k^{a}}\mathcal{A}_{\mathrm{\alpha }}^{b}\left( \mathbf{k}\right) \\
&=-i\sum\limits_{\gamma }\left[ r_{\alpha \gamma }^{a}\left( \mathbf{k}%
\right) r_{\gamma \alpha }^{b}\left( \mathbf{k}\right) -r_{\alpha \gamma
}^{b}\left( \mathbf{k}\right) r_{\gamma \alpha }^{a}\left( \mathbf{k}\right) %
\right], \\
&r_{\alpha \beta ;b}^{a}\left( \mathbf{k}\right) -r_{\alpha \beta
;a}^{b}\left( \mathbf{k}\right) \\
&=-i\sum\limits_{\gamma }\left[ r_{\alpha \gamma }^{a}\left( \mathbf{k}%
\right) r_{\gamma \beta }^{b}\left( \mathbf{k}\right) -r_{\alpha \gamma
}^{b}\left( \mathbf{k}\right) r_{\gamma \beta }^{a}\left( \mathbf{k}\right) %
\right], \\
&r_{\alpha \beta ;b}^{a}\left( \mathbf{k}\right) =\partial _{k^{b}}r_{\alpha
\beta }^{a}-i\left[ \mathcal{A}_{\mathrm{\alpha }}^{b}\left( \mathbf{k}%
\right) -\mathcal{A}_{\mathrm{\beta }}^{b}\left( \mathbf{k}\right) \right]
r_{\alpha \beta }^{a},
\end{aligned}
\end{equation*}%
we find%
\begin{equation}
j^{a}\left( t\right) =j_{\mathrm{intra}}^{a}\left( t\right) +j_{\mathrm{inter%
}}^{a}\left( t\right) +j_{\mathrm{top}}^{a}\left( t\right),   \label{tcur2}
\end{equation}%
where 
\begin{equation}
j_{\mathrm{intra}}^{a}\left( t\right) =-\sum\limits_{n\mathbf{k}}\partial
_{k^{a}}\mathcal{E}_{n}(\mathbf{k)}\varrho _{nn}(\mathbf{k},t)
\label{jintra}
\end{equation}%
is the intraband current,%
\begin{equation}
\begin{aligned}
&j_{\mathrm{inter}}^{a}\left( t\right) =-\sum\limits_{n\neq m}\sum\limits_{%
\mathbf{k}}\left\vert r_{mn}^{a}\left( \mathbf{k}\right) \right\vert
e^{i\left( \Phi _{nm}^{x}-\Phi _{nm}^{a}\right) } \\
&\times \left[ \partial _{t}\varrho _{nm}(\mathbf{k},t)+\Gamma \varrho _{nm}(%
\mathbf{k},t)\right]   \label{jinter}
\end{aligned}
\end{equation}%
is the interband current and 
\begin{equation}
\begin{aligned}
&j_{\mathrm{top}}^{a}\left( t\right) =\sum\limits_{n}\sum\limits_{\mathbf{k}%
}\epsilon ^{axc}E(t)\mathcal{B}_{n}^{c}\left( \mathbf{k}\right) \varrho
_{nn}(\mathbf{k},t) \\
&-\sum\limits_{n\neq m}\sum\limits_{\mathbf{k}}\left( \partial
_{k^{a}}\left\vert r_{mn}^{x}\left( \mathbf{k}\right) \right\vert
+iR_{nm}^{a,x}\left\vert r_{mn}^{x}\left( \mathbf{k}\right) \right\vert
\right) E(t)\varrho _{nm}(\mathbf{k},t)  \label{jtop}
\end{aligned}
\end{equation}%
represents the topological part of the current. Here $\epsilon ^{abc}$ is
the Levi-Civita symbol and the summation over the repeated upper indices is
implied. Note that we include a damping term in Eq. (\ref{jinter}) as part
of the interband current. In the topological contribution, we consider the
effects of both Berry curvature and the shift vector on equal footing, as
they both significantly influence the HWM spectra.

\section{Generation of two-color high harmonics}

We further explore the nonlinear response of biased BGL considering
two-color high harmonic generation and wave mixing. The wave--particle
interaction will be characterized by the dimensionless parameters $\chi
_{1,2}$ $=E_{01,2}a/\hbar \omega _{1,2}$, which represent the work of the
waves electric fields on a lattice spacing ($a$) in the units of
corresponding photon energy $\hbar \omega _{1,2}$. Here we take one of the
frequencies ($\omega _{1}$) in the domain from near-infrared to the UV one,
and the other of the pump waves ($\omega _{2}$) in the mid infrared/THz
domain of frequencies. For the high frequency wave we assume $\chi _{1}<<1$,
meanwhile for the low frequency wave $\chi _{2}\sim 1$. For the durations of
the pulses we take $\mathcal{\tau }_{1}=\mathcal{\tau }_{2}=8\mathcal{T}_{2}$%
, where $\mathcal{T}_{2}\mathcal{\ }$is the period of the low frequency
wave. The relaxation rate is assumed to be $2\pi \hbar \Gamma =0.1\,\mathrm{%
eV}$ ($\Gamma ^{-1}=40\ \mathrm{fs}$). 
\begin{figure*}[tbp]
\includegraphics[width=0.95\textwidth]{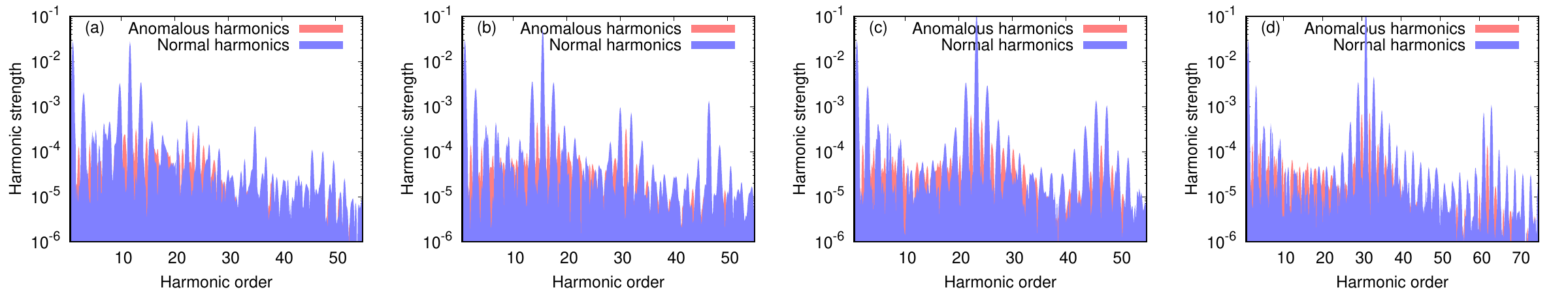}
\caption{HWM spectra in logarithmic scale. We assume $\hbar \protect\omega %
_{2}=0.1\,\mathrm{eV}$, $E_{02}=2\times 10^{6}\,\mathrm{V/cm}$ ($\protect%
\chi _{2}=0.5$), and $E_{01}=2.5\times 10^{6}\,\mathrm{V/cm.}$ From left to
right (a) $\hbar \protect\omega _{1}=1.17\,\mathrm{eV}$ ($\protect\chi %
_{1}=5.\,\allowbreak 3\times 10^{-2}$), (b) $\hbar \protect\omega _{1}=1.55\,%
\mathrm{eV}$ ($\protect\chi _{1}=4\times 10^{-2}$), (c) $\hbar \protect%
\omega _{1}=2.3\,\mathrm{eV}$ ($\protect\chi _{1}=2.7\times 10^{-2}$), and
(d) $\hbar \protect\omega _{1}=3.1\,\mathrm{eV}$ ($\protect\chi _{1}=2\times
10^{-2}$).}
\end{figure*}
By numerically solving the structure-gauge invariant semiconductor Bloch
equations (\ref{SBE1}) and (\ref{SBE2}) with a BZ sampled by $10^{5}$ k-points, we calculate
the time-dependent current using Eqs. (\ref{tcurinv}), (\ref{jintra}), (\ref%
{jinter}), and (\ref{jtop}). The HHG spectral yield is proportional
to the spectral intensity of the current, expressed as $\propto \omega
^{2}\sum_{a=x,y}\left\vert j^{a}\left( \omega \right) \right\vert ^{2}$,
where $j^{a}\left( \omega \right) $ is the Fourier transform of the $a$-th
component of the time-dependent current $j^{a}\left( \omega \right)
=\int_{-\infty }^{\infty }j^{a}\left( t\right) e^{i\omega t}dt$. Thus, for
the polarization-resolved high-order harmonic spectrum, we calculate the
quantity: $s^{a}\left( \omega \right) =\omega \left\vert j^{a}\left( \omega
\right) \right\vert $.

Since inversion symmetry is broken in biased BLG, there are no restrictions
on the total number of involved photons. As a result, the output radiation
can emerge at frequencies $n_{1}\omega _{1}\pm n_{2}\omega _{2}$, where $%
n_{1,2}=0,1,2,...$. However, to analyze the contribution of the topological
part of the current, we consider pump waves polarized along the K-G-K'
direction. Due to degenerecy of $K$ and $K^{\prime }$ valleys and mirror
symmetry of BZ the following selection rules for HWM apply: for HWM
generated along the laser polarization direction (referred to as normal
harmonics), $n_{1}\pm n_{2}$ is odd, while output radiation emerging along
the perpendicular direction (refered to as anomalous harmonics) has $%
n_{1}\pm n_{2}$ even. These selection rules can be understood through a
semi-classical analysis of electronic motion in conjunction with the
three-fold crystal symmetry \cite{liu2017high}.

In Fig. 5, we present the polarization-resolved HWM spectra for BLG in the
strong-field regime, shown on a logarithmic scale for different frequencies
of the high-frequency pump wave. The harmonic order is defined in units of
low frequency as $\omega /\omega _{2}$. The presence or absence of specific
harmonics with particular polarizations is consistent with the selection
rules. For the normal (anomalous) harmonics, the initial peaks in the
spectra correspond to $n_{1}=0$ and $n_{2}$ odd (even), followed by the main
peak at $n_{1}=1$ and $n_{2}$ even (odd), and then $n_{1}=2$ and $n_{2}$ odd
(even). Between these peaks, there are plateau regions where the strength of
the anomalous harmonics is comparable to that of the normal harmonics. This
plateau becomes more extended as the frequency of the first wave increases.

Before delving deeper into the numerical results, let us first perform an
approximate analytical analysis to gain insight into the numerical findings.
In Bernal-stacked BLG, the strong interlayer coupling results in non-zero
transition matrix elements for all pairs of bands. Additionally, the shift
vector is non-zero for all transitions. As a result, an electron excited by
the high-frequency wave into any conduction band can, after being
accelerated by the low-frequency wave, recombine with a hole in any of the
valence bands. These multiple transition channels can interfere, leading to
the complex HWM spectra observed. Consequently, the underlying dynamics
become intricate. However, when the transitions between one specific valence
band and one conduction band dominate, it becomes possible to derive
qualitative insights through an approximate analytical solution \cite{vampa2015semiclassical,li2019phase,uzan2020attosecond,yue2021expanded,avetissian2022efficient}. To achieve
this, we omit the triple phase product term in Eq. (\ref{SBE2}) and
consider only the transitions between valence and conduction bands without relaxation. Since
the population difference $\delta N_{nm}=N_{n}(\mathbf{k}%
,t)-N_{m}(\mathbf{k},t)$ varies slowly on the time scale of the
waves' periods, the formal solution to Eq. (\ref{SBE2}) for the interband
coherence can be written as:%
\begin{equation}
\begin{aligned}
&\varrho _{nm}(\boldsymbol{\kappa}(t),t)=i\int\limits_{-\infty }^{t}dt^{\prime
}\left\vert r_{nm}^{x}(\boldsymbol{\kappa }\left(t^{\prime }\right)
)\right\vert E(t^{\prime })\delta N_{nm}\left( \boldsymbol{\kappa }%
\left(t^{\prime }\right) \right) \\
& \times \exp \left[ -i\int\limits_{t^{\prime }}^{t}d\eta \left[ \mathcal{E}%
_{nm}\left( \boldsymbol{\kappa }\left(\eta \right) \right) +E(\eta
)R_{nm}^{x,x}\left( \boldsymbol{\kappa }\left(\eta \right) \right) \right] %
\right] ,  \label{ronm}
\end{aligned}
\end{equation}%
where $\boldsymbol{\kappa }(t) =\mathbf{k} +\mathbf{A}(t) $ and $\mathbf{A}\left(
t\right) =-\int\limits_{-\infty }^{t}E(\eta )d\eta $ is the vector
potential. In this approximation, the second term in Eq. (\ref{tcurinv}) dominates and represents the non-intraband current. This component also incorporates the topological contribution to the current and can be defined using the obtained solution (\ref{ronm}). Consequently, the non-intraband current can be written as follows:
\begin{equation}
    \begin{aligned}
        j^{a}\left( t\right) &=\frac{2}{(2\pi )^2}\int\limits_{\overline{BZ}}d\mathbf{k}\int\limits_{-\infty }^{t}dt^{\prime
}\sum\limits_{n\neq m}C_{nm}\left( \mathbf{k,}%
t^{\prime }\right)  \\
        &\times \exp \left[ -iS_{nm}^{a}\left( \mathbf{k,}t,t^{\prime
}\right) \right]A_{nm}\left( \mathbf{k,}t\right) +c.c.,  \label{jtt} 
    \end{aligned}
\end{equation}
where
\begin{equation*}
C_{nm}\left( \mathbf{k,}t^{\prime }\right) =\left\vert r_{nm}^{x}(\boldsymbol{%
\kappa }\left( t^{\prime }\right) )\right\vert E(t^{\prime })\delta N_{nm}\left( \boldsymbol{\kappa }\left( t^{\prime }\right) \right)
\end{equation*}%
is the electron-hole creation amplitude, 
\begin{equation*}
A_{nm}\left( \mathbf{k,}t\right) =\left\vert r_{mn}^{a}(\boldsymbol{\kappa }%
\left( t\right) )\right\vert \mathcal{E}_{mn}\left( \boldsymbol{\kappa }\left(
t\right) \right)
\end{equation*}%
is the electron-hole recombination amplitude and 
\begin{equation}
\begin{aligned}
S_{nm}^{a}\left( \mathbf{k,}t,t^{\prime }\right) &=\int\limits_{t^{\prime
}}^{t}d\eta \left[ \mathcal{E}_{nm}\left( \boldsymbol{\kappa }\left( \eta
\right) \right) +E(\eta )R_{nm}^{x,x}\left( \boldsymbol{\kappa }\left( \eta
\right) \right) \right] \\
&+\Phi _{nm}^{a}(\boldsymbol{\kappa }\left( t\right))-\Phi _{nm}^{x}(\boldsymbol{\kappa }\left( t\right))  \label{action}
 \end{aligned}
\end{equation}%
is the action which differs from the classical action $\int\limits_{t^{%
\prime }}^{t}\mathcal{E}_{nm}\left( \boldsymbol{\kappa }\left( \eta \right)
\right) d\eta $ by the terms which are of topological nature. In Eq. (\ref{jtt}) the BZ
is also shifted to $\overline{BZ}=BZ-\mathbf{A}(t)$. The expresion
for the non-intraband current have a transparent physical interpretation in
analogy with the atomic three-step model: electron-hole creation at $%
t^{\prime }$ with amplitude $C_{nm}\left( \mathbf{k,}t^{\prime }\right) $,
then propagation in the BZ, which is defined by the action (\ref{action}). Finally,
the electron-hole pair annihilates at $t$ with the amplitude $A_{nm}\left( 
\mathbf{k,}t\right) $. 

For the Fourier transform of the non-intraband current, we will have%
\begin{equation}
\begin{aligned}
&j^{a}\left( \omega \right) =\frac{2}{(2\pi )^2}\int\limits_{\overline{BZ}}d\mathbf{k}\int\limits_{-\infty }^{\infty
}dt\int\limits_{-\infty }^{t}dt^{\prime }\sum\limits_{n\neq m}C_{nm}\left( \mathbf{k,}t^{\prime }\right) \\
&\times \exp \left[
-iS_{nm}^{a}\left( \mathbf{k,}t,t^{\prime }\right) +i\omega t\right]
A_{nm}\left( \mathbf{k,}t\right) +\textrm{c.c.}(\omega \rightarrow -\omega ). \label{FTJ}
\end{aligned}
\end{equation}
As seen from this formula, HHG can be driven by the nonlinearity of the
transition velocities, which define the annihilation amplitude $A_{nm}\left( 
\mathbf{k,}t\right) $, as well as by the fast oscillatory component
originating from the action in Eq. (\ref{action}). Assuming that the
dominant mechanism driving HHG is the latter, we can further develop the
collision model by evaluating the integrals in Eq. (\ref{FTJ}) using the
saddle-point method. Given that $\omega _{1}>>$ $\omega _{2}$, it is 
necessary to account for the rapid oscillations induced by the
high-frequency component of the field. Note that at $\chi _{2}>>\chi _{1}$ in the
propogation part the main role playes low frequency wave. By employing a Taylor expansion of
the creation amplitude over the high-frequency field, expressed as $%
C_{nm}\left( \mathbf{k,}t^{\prime }\right) =\sum_{n_{1}}\Lambda _{nm}\left( 
\mathbf{k,}t^{\prime },n_{1}\right)\chi _{1}^{n_{1}}\exp (in_{1}\omega _{1}t^{\prime
})$, it becomes apparent that the resulting saddle-point
conditions should be imposed on the phase term $\Delta \left( \mathbf{k,}%
t,t^{\prime }\right) =S_{nm}^{a}\left( \mathbf{k,}t,t^{\prime }\right)
-\omega t+n_{1}\omega _{1}t^{\prime }$. Taking into account Eq. (\ref{Berry2}%
) and the relation $\partial _{t}\kappa _{x}\left( t\right) =-E(t)$, we
derive the saddle-point conditions from the equations $\partial _{t^{\prime
}}\Delta \left( \mathbf{k,}t,t^{\prime }\right) =0$, $\partial _{t}\Delta
\left( \mathbf{k,}t,t^{\prime }\right) =0$, and $\partial _{\mathbf{k}%
}\Delta \left( \mathbf{k,}t,t^{\prime }\right) =0$. The resulting
saddle-point conditions are given by:

\begin{equation}
\mathcal{E}_{nm}\left( \boldsymbol{\kappa }\left( t^{\prime }\right) \right)
+E(t^{\prime })R_{nm}^{x,x}\left( \boldsymbol{\kappa }\left( t^{\prime }\right)
\right) =n_{1}\omega _{1},  \label{sd1}
\end{equation}%
\begin{equation}
\mathcal{E}_{nm}\left( \boldsymbol{\kappa }\left( t\right) \right)
+E(t)R_{nm}^{x,a}\left( \boldsymbol{\kappa }\left( t\right) \right) =\omega ,
\label{sd2}
\end{equation}%
\begin{equation}
\delta r^{b}=R_{nm}^{b,a}(\boldsymbol{\kappa }\left( t\right) )-R_{nm}^{b,x}\left( 
\boldsymbol{\kappa }\left( t^{\prime }\right) \right) ,  \label{sd3}
\end{equation}%
with the electron-hole separation vector 
\begin{equation}
\delta r^{b}=\int\limits_{t^{\prime }}^{t}d\eta \left[ \mathrm{v}%
_{n}^{b}\left( \boldsymbol{\kappa }\left( \eta \right) \right) -\mathrm{v}%
_{m}^{b}\left( \boldsymbol{\kappa }\left( \eta \right) \right) \right] ,
\label{sd4}
\end{equation}%
and group velocity%
\begin{equation}
\mathrm{v}_{n}^{b}=\partial _{k^{b}}\mathcal{E}_{n}(\mathbf{k)+}\epsilon
^{bxc}E(t)\mathcal{B}_{n}^{c}\left( \mathbf{k}\right).  \label{sd5}
\end{equation}
Equations (\ref{sd1}) - (\ref{sd3}) can be interpreted as follows. The first
equation defines the birth time, or the moment at which the electron-hole
pair is formed. It also indicates that the electron-hole pair is created
with an initial momentum determined by the absorption of high-energy
photons. The peaks observed in Fig. 5, near the multiples of the high
frequency $n_{1}\omega _{1}$, highlight the relevance of this condition. The
second equation (\ref{sd2}) represents the conservation of energy:
electron-hole recombination results in the emission of a photon whose energy
equals the band gap at the recombination crystal momentum, plus an
additional contribution from the dipole interaction energy arising from the
shift vector. Since the acceleration of the carriers is primarily driven by
the low-frequency wave, sidebands are observed near the main peaks in Fig.
5. The third equation (\ref{sd3}) defines the relative spatial distance
between the electron and hole at the time of recombination. In the
structure-gauge invariant formalism, these saddle-point equations explicitly 
include two key topological quantities and are equivalent to those derived
in Refs. \cite{li2019phase,yue2021expanded}, where the shift vector was not introduced. These equations
differ significantly from their counterparts in materials where both
time-reversal and inversion symmetries are preserved \cite{vampa2015semiclassical}. As seen, Eqs. (\ref%
{sd1}) and (\ref{sd2}) feature an additional term for the dipole interaction
energy originating from the shift vector. For electron-hole trajectories
described by Eqs. (\ref{sd4}) and (\ref{sd5}), we encounter the anomalous
velocity term, which is governed by the Berry curvature. This results in
electrons and holes exhibiting velocities perpendicular to the pump wave's
electric field. Additionally, Eq. (\ref{sd3}) introduces a shift in the
recollision condition, which distinguishes between normal and anomalous
harmonics. In materials possessing both time-reversal and inversion
symmetries, anomalous harmonics are absent, and Eq. (\ref{sd3}) simplifies
to $\delta r^{b}=0$, implying that the total distance traveled by the
electron equals the total distance traveled by the hole. If the electron and
hole originate from the same position, this condition ensures that high
harmonics are emitted only when the electron and hole precisely re-encounter
each other. In our system, however, electron-hole creation occurs with a
displacement dictated by the shift vector. Equation (\ref{sd3}) establishes
that the total difference in distance traveled by the electron and hole
corresponds to the difference in the shift vector at their points of
creation and annihilation. 

Once the solutions  ($\overline{t},\overline{t^{\prime }},\overline{\mathbf{k%
}}$) to the saddle-point equations (\ref{sd1}) - (\ref{sd3}) are determined,
the frequency-resolved non-intraband current can be expressed as%
\begin{equation}
\begin{aligned}
&j^{a}\left( \omega \right) \propto \sum\limits_{n\neq m}\sum\limits_{n_{1}}\sum\limits_{%
\overline{t},\overline{t^{\prime }},\overline{\mathbf{k}}}\frac{\chi _{1}^{n_{1}}\Lambda_{nm}\left( 
\overline{\mathbf{k}}\mathbf{,}\overline{t^{\prime }}, n_{1}\right) A_{nm}\left( 
\overline{\mathbf{k}}\mathbf{,}\overline{t}\right) }{\sqrt{\left\vert \det %
\left[ \partial ^{2}S_{nm}^{a}\left( \overline{\mathbf{k}}\mathbf{,}%
\overline{t},\overline{t^{\prime }}\right) \right] \right\vert }} \\
&\times\exp \left[
-iS_{nm}^{a}\left( \overline{\mathbf{k}}\mathbf{,}\overline{t},\overline{%
t^{\prime }}\right) +i\omega \overline{t}+in_{1}\omega_{1} \overline{t^{\prime }}\right] +\textrm{c.c.}(\omega \rightarrow
-\omega), \label{hes}
\end{aligned}
\end{equation}
where $\partial ^{2}S_{nm}^{a}$ is the Hessian matrix of the function $S_{nm}^{a}$ at the stationary points. Solving the
saddle-point equations (\ref{sd1}) - (\ref{sd3}) in their full quantum form
is a monumental task and falls outside the scope of this paper.
Nevertheless, a significant amount of physical insight can be gained from a
qualitative solution, which will be explored subsequently.

As is well-known, graphene exhibits a peak in linear absorption when a
photon is in resonance with the van Hove singularity (vHS) at the M point of
the BZ. For BLG we will have several peaks due to non-zero transition matrix
elements for all pairs of bands. Therefore, it is of particular interest to
explore the case where the high-frequency driving wave is in one-photon
resonance with the vHS. From Figure 2, it can be seen that the one-photon
resonant frequency for the vHS for the transition $1\rightarrow 2$ is
approximately $\omega _{1}\simeq 4-4.2\,\mathrm{eV}$\textrm{.} Figures 6(a)
and 6(c) show the HHG spectra for anomalous and normal harmonics at both
resonant and non-resonant frequencies. As expected, the resonant frequency
leads to a significant enhancement in the nonlinear optical response of BLG.
To ensure resonant enhancement, we performed calculations for different 
frequencies $\omega _{1}$ within the range $3.5\div 5.0\,\mathrm{eV}
$, keeping the electric field strength fixed. We summed the harmonic strength 
over sidebands near the main frequency to obtain the total harmonic yield near the resonance:
$\overline{s^{a}}(\omega _{1}) =%
\sum\limits_{n=-10}^{10}s^{a}\left( \omega _{1}+n\omega _{2}\right) $. The resulting curves for 
anomalous and normal harmonics are shown in Fig. 6(b) and Fig. 6(d), respectively. As evident 
from these figures, there is indeed a resonant enhancement.
Considering the presence of multiple transition channels (see Figs. 10 and 11) with 
different resonant frequencies and corresponding yields, the resonance width is 
relatively broad. For normal harmonics, the resonance peak 
is blueshifted compared to anomalous harmonics. This blueshift occurs because 
transitions $0\rightarrow 2$ and $1\rightarrow 3$ dominate in the case of normal harmonics.

\begin{figure}[tbp]
\includegraphics[width=0.5\textwidth]{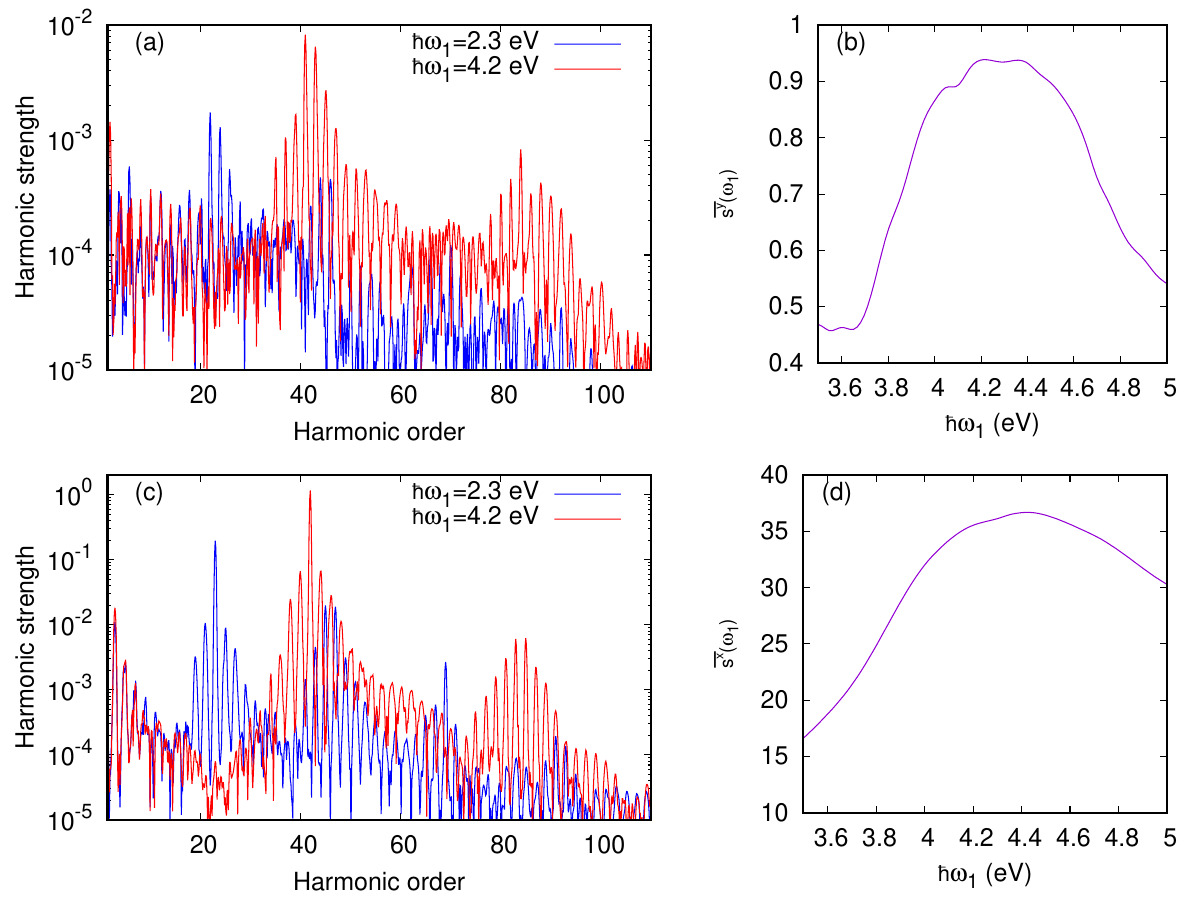}
\caption{HWM spectra for both non-resonant and one-photon resonant frequencies: (a)
anomalous harmonics and (c) normal harmonics. We assume $\hbar \protect%
\omega _{2}=0.1\,\mathrm{eV}$, $E_{02}=5\times 10^{6}\,\,\mathrm{V/cm}$ ($%
\protect\chi _{2}=\allowbreak 1.\,\allowbreak 25$), and $E_{01}=5\times
10^{6}\,\mathrm{V/cm}$ ($\protect\chi _{1}=5.\,\allowbreak 3\times 10^{-2}$; 
$\allowbreak 3\times 10^{-2}$). The dependence of the harmonic strength summed 
over sidebands on frequency $\omega _{1}$ is depicted in (b) and (d).}
\end{figure}
\begin{figure}[tbp]
\includegraphics[width=0.5\textwidth]{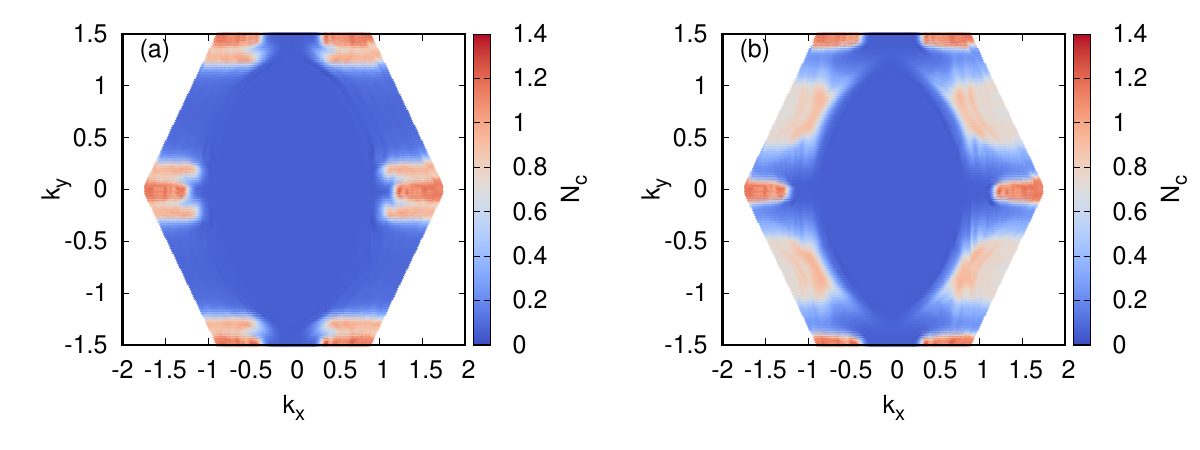}
\caption{Conduction band population following interaction with a pump field
at one-photon resonant and non-resonant frequencies, corresponding to the interaction
parameters in Fig. 6: (a) corresponds to $\hbar \protect\omega _{1}=2.3\,%
\mathrm{eV}$ and (b) corresponds to $\hbar \protect\omega _{1}=4.2\,\mathrm{%
eV}$.}
\end{figure}
\begin{figure}[tbp]
\includegraphics[width=0.46\textwidth]{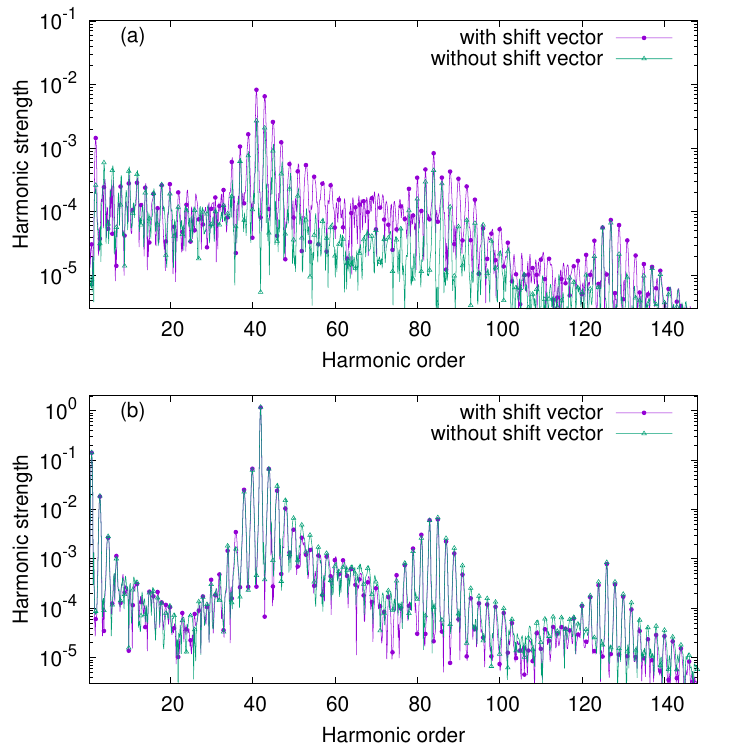}
\caption{The role of the shift vector in the HWM spectra for (a) anomalous
harmonics and (b) normal harmonics. The parameters used are $\hbar \protect%
\omega _{2}=0.1\ \mathrm{eV}$, $E_{02}=5\times 10^{6}\ \mathrm{V/cm}$ ($%
\protect\chi _{2}=\allowbreak 1.25$), and $\hbar \protect%
\omega _{1}=4.2\ \mathrm{eV}$, $E_{01}=5\times 10^{6}\ \mathrm{%
V/cm}$ ($\protect\chi _{1}=3\times 10^{-2}$).}
\end{figure}
Figure 7 illustrates the conduction band population $N_{c}(\mathbf{%
k)}=N_{2}(\mathbf{k)}+N_{3}(\mathbf{k)}$ after
interaction with a pump field at resonant and non-resonant frequencies. At
the resonant frequency, a substantial population near the $M$ points is
observed, where the energy dispersion curves flatten (as shown in Figure 2).
The results presented in Fig. 6 can be further explained through the
saddle-point solutions. Consider the
case $n_{1}=1$ in Eq. (\ref{sd1}) when the high-frequency driving wave is in
one-photon resonance with the vHS. Near the vHS, the bands become flat,
meaning electron-hole pairs are created with approximately zero velocities: $%
\partial _{k^{b}}\mathcal{E}_{n}(\mathbf{k})\simeq 0$. In this scenario,
there is no additional separation caused by their opposite initial
velocities. This condition is particularly favorable for the re-encounter of
electron-hole pairs in a linearly polarized wave field. Furthermore, the
Hessian determinant is proportional to $\partial _{t}\partial _{k^{b}}S\left( \mathbf{k,}%
t,t^{\prime }\right)$ \cite{uzan2020attosecond} and tends to zero at the vHS. This implies that for
small values of $\partial _{k^{b}}\mathcal{E}_{n}(\mathbf{k)}$, the harmonic
yield at the corresponding harmonic energy is significantly enhanced. This phenomenon, referred to as spectral 
singularities in \cite{uzan2020attosecond}, highlights the unique interplay between vHS 
and the nonlinear optical response. Additionally, as evident from 
Eq. (\ref{sd1}), electron-hole pairs can
also be created near the vHS through a two-photon resonant process. Although
the probability of such excitations is smaller for $\chi _{1}<<1$
due to the prefactor $\chi _{1}^{n_{1}}$ in Eq. (\ref{hes}), this
process can still be enhanced due to contributions from the Hessian determinant. Such
behavior is observed in Fig. 6(c), particularly for harmonics H45 and H47,
when the excitation energy is  $\hbar \omega _{2}=2.3\ \mathrm{eV}$. Here, the
two-photon resonance condition aligns closely with the vHS (see Fig. 6(d)),
resulting in enhanced harmonic generation.
\begin{figure}[tbp]
\includegraphics[width=0.46\textwidth]{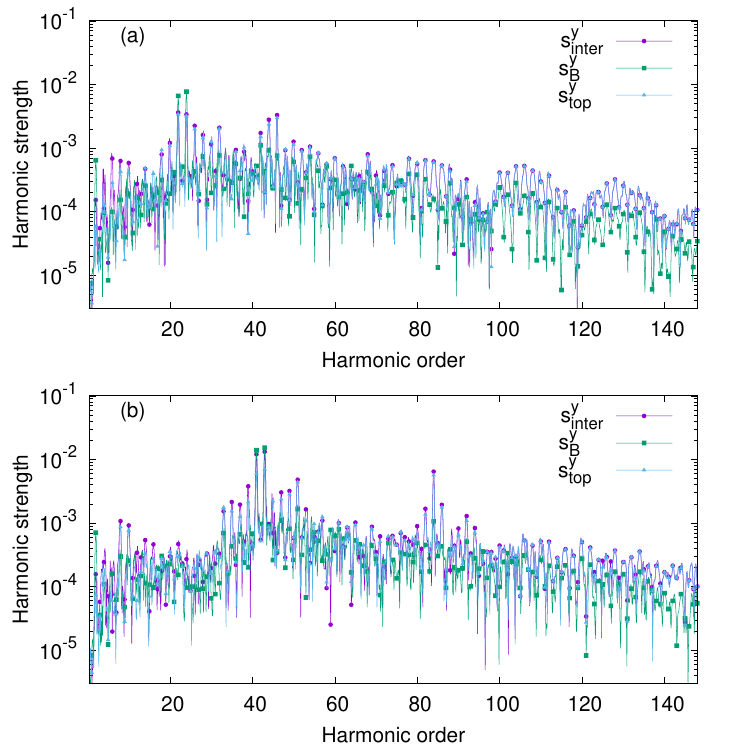}
\caption{The contribution of various components of the current to the HWM
spectra for anomalous harmonics, corresponding to the interaction parameters
in Fig. 6: (a) corresponds to $\hbar \protect\omega _{1}=2.3\ \mathrm{eV}$,
and (b) corresponds to $\hbar \protect\omega _{1}=4.2\ \mathrm{eV}$.}
\end{figure}
\begin{figure}[tbp]
\includegraphics[width=0.46\textwidth]{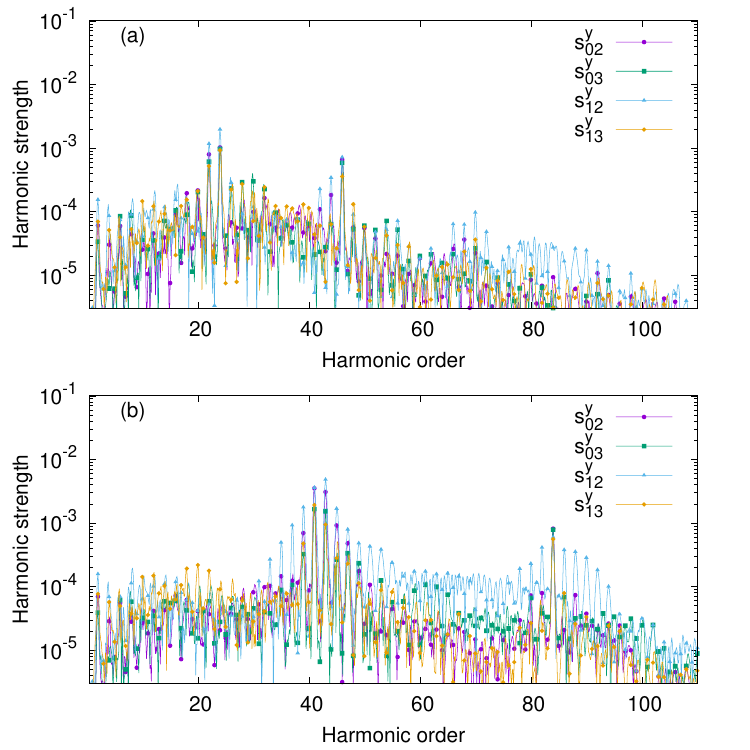}
\caption{The contribution of various interband transitions to the HWM
spectra for anomalous harmonics, corresponding to the interaction parameters
in Fig. 6: (a) corresponds to $\hbar \protect\omega _{1}=2.3\ \mathrm{eV}$,
and (b) corresponds to $\hbar \protect\omega _{1}=4.2\ \mathrm{eV}$.}
\end{figure}

\begin{figure}[tbp]
\includegraphics[width=0.46\textwidth]{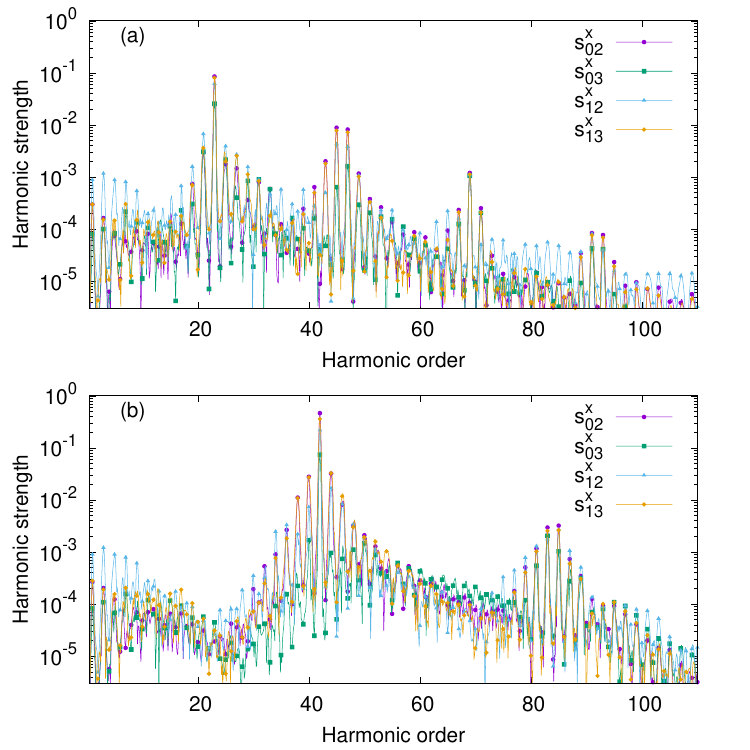}
\caption{Same as in Fig. 10, but for normal harmonics.}
\end{figure}
\begin{figure}[tbp]
\includegraphics[width=0.46\textwidth]{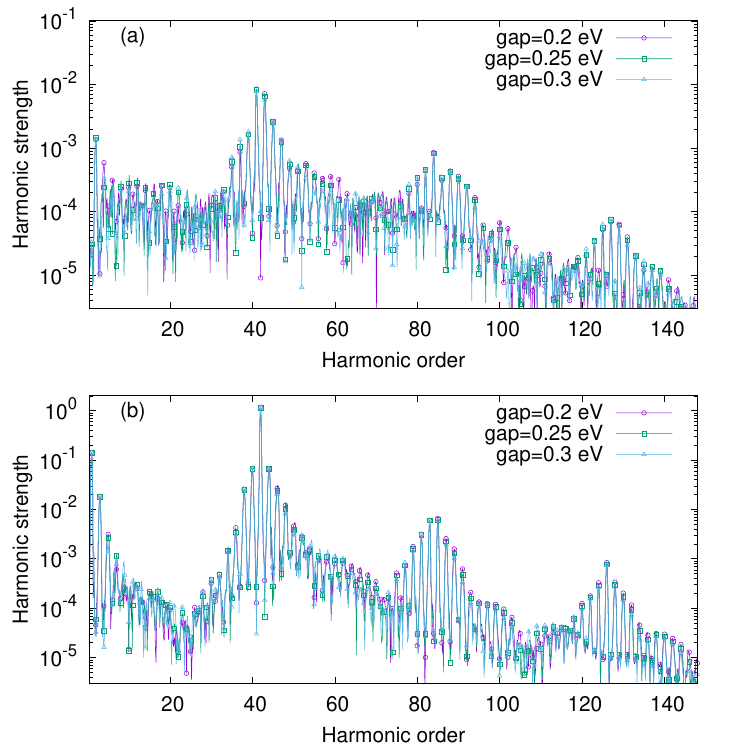}
\caption{The role of the energy gap in the HWM spectra for (a) anomalous
harmonics and (b) normal harmonics, corresponding to the interaction
parameters in Fig. 10.}
\end{figure}

A deeper understanding of the HWM mechanism can be achieved by analyzing the
role of the shift vector in the HWM spectra. We solved the Bloch equations
both with and without incorporating the shift vector $%
R_{nm}^{x,x}$ and compared the resulting spectra in Fig. 8. Unlike normal
harmonics, where the shift vector plays only a quantitative role, it
significantly alters and reshapes the spectra for anomalous harmonics. This
demonstrates that the topological component of the current, influenced by
the shift vector, has a substantial impact on the HWM spectra. These results can be clearly interpreted 
using the saddle-point solutions. When the shift vector component $R_{nm}^{x,x}$ is set to zero in Eq. (\ref{SBE2}), 
the saddle-point equation (\ref{sd3}) transforms into:
\begin{equation}
\delta r^{b}=R_{nm}^{b,a}(\boldsymbol{\kappa }\left( t\right) )-R_{nm}^{b,x}\left( 
\boldsymbol{\kappa }\left( t\right) \right)   \label{skhal}
\end{equation}
and the anomalous velocity term vanishes, meaning there is no force component in the y-direction. Near the vHS, electron-hole pairs are created with approximately zero velocities. Consequently, their motion becomes effectively constrained to one dimension along the x-axis, resulting in $\delta r^{y}=0$.
For anomalous harmonics ($a=y$), Eq. (\ref{skhal}) imposes a strict condition: $R_{nm}^{y,y}(\mathbf{k})-R_{nm}^{y,x}\left( \mathbf{k} \right) =0$, which must be satisfied irrespective of the trajectories. This condition is only met at specific points in k-space, leading to a suppression of HHG for anomalous harmonics, as shown in Fig. 8(a). This phenomenon provides a clear manifestation of topological effects, specifically highlighting the roles of the Berry curvature and the shift vector in shaping the nonlinear electrodynamic response of BLG. For normal harmonics ($a=x$), Eq. (\ref{skhal}) simplifies to $\delta r^{b}=0$. This condition ensures that high harmonics are emitted only when the electron and hole re-encounter each other without any displacement caused by the shift vector. As a result, a slight enhancement in the HHG spectra is observed when $R_{nm}^{x,x}$ is set to zero in Eq. (\ref{SBE2}), as shown in Fig. 8(b).

To further elucidate how different components of the current affect the HWM
spectrum of BLG, Fig. 9 shows the spectral contributions of the interband
current, the topological current, and the current influenced solely by the
Berry curvature. As evident from the figure, the topological and interband
currents contribute approximately equally. The harmonics where the Berry
curvature dominates are $n_{1}=0$, $n_{2}=2$, and $n_{1}=1$, $n_{2}=\pm 1$. 

Next, we examine how HWM depends on various interband transition channels.
In Bernal-stacked BLG, strong interlayer coupling results in non-zero
transition matrix elements for all pairs of bands. These transition channels
can interfere, leading to the observed HWM spectra. The contribution of
various interband transitions to the HWM spectra for anomalous and normal
harmonics is illustrated in Figs. 10 and 11, respectively. As shown, all
transitions contribute significantly to the spectra. For anomalous
harmonics, the transition from the top of the valence band to the bottom of
the conduction band is dominant. This transition, with its minimal energy
gap, ensures maximum Berry curvature and shift vector, which are crucial in
defining the topological part of the current. In contrast, for normal
harmonics, the contributions from various transitions are more balanced. 

Finally, let's consider the role of interlayer bias on the harmonic spectra.
In Fig. 12, we plot the HWM spectra for various interlayer biases. As seen,
normal harmonics are robust against changes in interlayer bias, as interband
harmonics dominate in this case. However, for anomalous harmonics, the
high-frequency part of the spectrum remains stable with respect to
interlayer bias changes, whereas the lower-frequency part is sensitive to
such changes. This sensitivity arises because these harmonics are influenced
by the Berry curvature and shift vector, which are sensitive to the gap
magnitude near the Dirac points.

\section{Conclusion}
We have presented the structure-gauge invariant microscopic theory of the nonlinear interaction of a BLG with broken inversion symmetry and strong two-color laser fields. The developed theory accounts for the full BZ of the hexagonal 2D nanostructure. For concreteness, we have considered the nonlinear excitation of biased BLG by two-color pump waves toward high-order wave mixing. Specifically, we examined the case where the high-frequency pump wave generates electron-hole pairs with large initial momentum away from the $K$ points. The low-frequency wave then accelerates these
electron-hole pairs, and after recombination, two-color high harmonic generation and wave mixing are observed. We observed harmonics generated along the polarization direction of the lasers, characterized by an odd number of involved photons, as well as intense anomalous harmonics along the perpendicular direction, where an even number of photons are mixed. The intensities of anomalous harmonics are strongly dependent on the Berry curvature and shift vector. Particular attention was given to the scenario where a high-frequency driving wave is in one-photon resonance with the vHS
at the $M$ point of the BZ. In this case, a considerable enhancement is observed in the HWM spectrum compared to the nonresonant case. The HWM was also analized with a quasiclassical method including
the classical trajectory of electron-hole pairs, where Berry curvature and the shift vector significantly influence the saddle-point equations. The obtained results demonstrate that the biased BLG nanostructure can serve as an effective medium for high-order wave mixing and harmonic generation under multiphoton excitation by two-color strong laser pulses.

\begin{acknowledgments}
The work was supported by the Science Committee of Republic of
Armenia, project No. 21AG-1C014.
\end{acknowledgments}

\bibliographystyle{apsrev4-2}
\bibliography{liter}

\end{document}